\documentclass[aps,prl,twocolumn,nofootinbib,superscriptaddress]{revtex4-2}

\usepackage{amsmath,amssymb,amsfonts}
\usepackage{bm}
\usepackage{hyperref}
\hypersetup{colorlinks = true,
  urlcolor = blue,
  linkcolor = blue,
  citecolor = blue,
  pdftitle = {Emergent supersymmetry in a time-space inverted quantum mechanics}
}
\usepackage{physics}
\usepackage{bbm}

\usepackage{xcolor,cancel}
\usepackage[normalem]{ulem}
\usepackage{dsfont}

\begin{document}

\title{Emergent Supersymmetry in Space–Time Inverted Quantum Mechanics}

\newcommand{\UFPR}{Departamento de Física, Universidade Federal do Paran\'a, 81531-980 Curitiba, Paran{\'a}, Brazil}
\newcommand{\CICTI}{Centro Interdisciplinar de Ciência, Tecnologia e Inovação (CICTI), Núcleo de Modelagem e Computação Científica (NMCC), Laboratório  Chaomplexity, Universidade Federal do Paran\'a, 81531-980, Curitiba - PR, Brazil}
\newcommand{\MPIPKS}{Max-Planck Institute for the Physics of Complex Systems, Nöthnitzer Str. 38, 01187 Dresden, Germany}

\author{Marcus W.~Beims}
\email{mbeims@fisica.ufpr.br}
\affiliation{\UFPR}
\affiliation{\CICTI}

\author{Arlans J.~S.~de Lara}
\email{arlans@pks.mpg.de}
\affiliation{\UFPR}
\affiliation{\CICTI}
\affiliation{\MPIPKS}

\date{\today}

\begin{abstract}
This Letter shows that a supersymmetric structure is inherent to the time-space–inverted (TSI) quantum mechanics (QM) framework, where the spatial evolution of states is generated by the operator $\hat{\mathcal{P}}^{\pm}(\hat{\mathcal{H}},\hat t;q)=\pm\sqrt{2m[\hat{\mathcal{H}}-\mathcal{\hat V}(q)]}$ [\href{https://doi.org/10.1103/PhysRevA.95.032133}{Phys. Rev. A. {\bf 95}, 032133 (2017)}],  named here \textit{Momentunian}, whose square-root structure that can be factorized. Such factorization leads directly to a supersymmetric algebra with supercharges and partner Hamiltonians. For the relativistic Momentunian the zero mode states are shown to be evanescent  states, \textit{independent} of the physical potential. Furthermore,  the existence of non-relativistic and relativistic Momentunian \textit{partners} is demonstrated, whose zero-mode states are no longer necessarily zero energies, but vanishing momenta states.  The natural emergence of the $1/2$-fractional time derivatives in the TSI QM, leads to supercharges which incorporate memory effects into the supersymmetric wave functions. Results indicate that supersymmetry emerges as a structural property of the TSI QM rather than being imposed phenomenologically.
\end{abstract}

\maketitle


\noindent
\textit{Introduction.}
The asymmetry between space and time in QM remains one of its most persistent conceptual problems. While spatial coordinates are represented by operators acting on a Hilbert space, time appears only as a parameter. This asymmetry is connected  to deep questions concerning the measurement of time, the status of energy-time uncertainty relations, and the absence of a self-adjoint time operator canonically conjugate to the Hamiltonian.  The problem of defining time in QM has been addressed from conceptual and operational definitions \cite{Razavy1967,LandauerButtiker1982,LandauerMartin1994,MugaBrouardMacas1995,GrotRovelliTate1996,GalaponCaballarBahague2004,Yearsley2011,GiovannettiLloydMaccone2015,Halliwell2015,GALAPON2018278,FloresGalaponTunneling2022,FloresGalaponTunneling2022}, and by investigating the emergence of time in quantum systems coupled to their environments \cite{Page1983,Rovelli1990,BriggsRost2000,BriggsRost2001,MarolfRovelli2002,Briggs2008,Gemsheim2023}. 
More recently, a space–time-symmetric (STS) formalism was proposed by Dias and Parisio \cite{DiasParisio2017}, where an extended Hilbert space is used, namely $\mathcal{H}_E = \mathcal{H}_X  \otimes \mathcal{H}_T$, where $\mathcal{H}_X$  is the usual Hilbert subspace with  time being a parameter and $\mathcal{H}_T$ is the additional Hilbert subspace which considers time as an operator and the position as a parameter. They  proposed that the momentum operator 
\begin{equation}
\label{PI}
    \hat{\cal P}^{\pm}(\hat{\mathcal{H}},\hat t;q)=\pm\sqrt{2m[\hat{\mathcal{H}}-\mathcal{\hat V}(q)]}, 
\end{equation}
generates the spatial evolution of states in $\mathcal{H}_T$. Here $\hat{\mathcal{H}}$ and $\mathcal{\hat V}(q)$ represent  respectively the total energy (canonical conjugated to the time operator) and potential operators in subspace  $\mathcal{H}_T$. The determination of tunneling and arrival times in this context have been discussed for specific physical situations \cite{DiasParisio2017,LaraBeims2024}, and a generalization of the non-relativistic STS to $3+1$ dimensions was studied \cite{Dias2025}.
 
The proposed STS formalism has a solid theoretical foundation, as it follows from the canonical quantization of an intriguing classical reinterpretation of variables, namely inverting the roles of space and time \cite{BeimsLara2024}. Such TSI formulation of classical mechanics (CM) leads to the \textit{Momentunian} $\mathcal{P}^{\pm}{}({\mathcal{H}}, t;q)=\pm\sqrt{2m[{\mathcal{H}}-\mathcal{V}(q)]} $, which, upon canonical quantization, reproduces exactly the operator (\ref{PI}).
It obeys the equation ($|\phi\rangle\in \mathcal{H}_T$)
\begin{equation}
\label{P}
    \hat{\mathcal{P}}^{\pm}(\hat{\mathcal{H}},\hat t;q)|\phi\rangle=-i\hbar\partial_q|\phi\rangle,
\end{equation}
which plays a role analogous to the Schrödinger equation (SE), but with position as the evolution parameter\footnote{The interpretation for this is that, whenever the Momentunian description is valid, it means that detectors are classical devices in which we know with certainty its position.}. The Momentunian assumes the role of generator of the spatial evolution, and while classical trajectories remain unchanged, QM is profoundly affected. 

From the supersymmetry perspective, Coleman and Mandula \cite{Coleman1967} demonstrated that unification of nature's symmetries was impossible using Lie algebras, leading to the development of superalgebras that define SUSY.  The SUSY formulation  by Witten \cite{Witten1981} evolved from a simplified toy model into a robust framework for creating Hamiltonian pairs with closely related eigenspectra, leading to supersymmetric quantum mechanics (SUSYQM) \cite{CooperKhareSukhatme1995,CooperKhareSukhatme2001}. The central idea consists in factorizing a 1D Hamiltonian into products of first-order differential operators, generating pairs of isospectral Hamiltonians connected by supercharges
\begin{equation}
\label{Qs}
    \hat Q=\begin{pmatrix}0&0\\ \hat A&0\end{pmatrix},\quad \mbox{and}\quad
    \hat Q^{\#}=\begin{pmatrix}0& \hat B\\0&0\end{pmatrix},
\end{equation}
with operators
\begin{align}
\label{As}
    \hat A = {-}\frac{\hbar}{\sqrt{2m}}\frac{d}{dq} {-}W(q),\quad
    \hat B = \frac{\hbar}{\sqrt{2m}}\frac{d}{dq} {-}W(q),
\end{align}
(in this case ${\#=\dagger}$ meaning that $\hat B=\hat A^{\dagger}$), named Hamiltonian partners (Left and Right sectors):
\begin{equation}
\label{Hsusy}
    \hat H_{\mbox{\tiny L,R}}=-\frac{\hbar^2}{2m}\partial^2_q+  W^2(q)\pm \frac{\hbar}{\sqrt{2m}} W'(q).
\end{equation}
We use the convention where the lower sign corresponds to the subindex to the left. The prime means derivative in respect to $q$. Generally, the superpotential $W(q)$ is an auxiliary function whose square and derivative generates the physical potentials.

Traditionally, SUSYQM has been formulated in the spatial domain, where the SE involves second-order derivatives with respect to position. This structure naturally enables the factorization of the Hamiltonian in terms of first-order spatial operators. On the other hand, the fact that in standard QM the SE has a first-order time derivative, a time-factorization procedure has not been included.  However, second-order time derivatives, which appear in wave equations such as Maxwell's equations, were used in the context of optics \cite{GarciaMeca2020}, enabling a direct transposition of the SUSYQM factorization procedure to the temporal domain, named T-SUSY.

In this work, we explore a fundamentally different realization of SUSYQM, which we refer to as $\sqrt{\mbox{T}}$-SUSY due to the square-root appearing in $\hat{\mathcal{P}}^{\pm}(\hat{\mathcal{H}},\hat t;q)$. We demonstrate that the TSI QM in subspace $\mathcal{H}_T$  naturally leads to SUSY with spatial operators. 
For time-independent potentials in the non-relativistic case, the superpotential is uniquely identified with the physical potential through $W(q)=\sqrt{\mathcal{V}(q)}$. For the relativistic Momentunian, the zero mode states are shown to be evanescent  states, \textit{independent} of the physical potential. Furthermore, we define fractional non-hermitian supercharges to obtain Momentunian partners. The zero modes states of the Momentunian partners are states with vanishing momenta, which corresponds to a constraint on the \textit{time evolution} of the wavefunctions. The $\sqrt{\mbox{T}}$-SUSY organizes the spectrum in momentum doublets rather than energy doublets.  The Momentunian partner construction naturally introduces $1/2$-fractional time derivatives at the algebraic level.

\noindent
\textit{Non-relativistic Momentunian.}
Assuming that $\mathcal{\hat H}$ and $\mathcal{\hat V}(q)$ are self-adjoint, the operators $\hat{\cal P}^{\pm}(\hat{\mathcal{H}},\hat t;q)$  have their domain restricted to the subspace where $\mathcal{\hat H}-\mathcal{\hat V}(q)\ge 0$, and define two self-adjoint operators with spectra \([0,\infty)\) and \((-\infty,0]\), respectively. A single self-adjoint operator with symmetric spectrum is obtained by introducing a two-component structure and defining 
$\mathcal{\hat{P}}(\hat{\mathcal{H}},\hat t;q) = \sigma_z \sqrt{2m\left[\mathcal{\hat H} - \mathcal{\hat V}(q)\right]},$
whose spectrum spans the entire real line. Here $\sigma_z=\mbox{diag}(1,-1)$. The time operator $\hat{t}$ is canonically conjugate to minus the energy operator  ($-\mathcal{\hat H}$), thus $\hat t|t\rangle=t|t\rangle$ and $[\hat t,\mathcal{\hat H}]=-i\hbar$. The set of eigenkets $\{ |t\rangle\}$ satisy the identity $\int_{-\infty}^{\infty} dt\, |t\rangle\langle t|=\mathbb{I}$ and $\Delta\hat t\Delta\mathcal{\hat H}\ge\hbar/2$. Applying Dirac’s factorization procedure, Eq.~(\ref{P}) transforms to \cite{BeimsLara2024} (using $\langle t|\sqrt{\mathcal{\hat H}}|\phi(q)\rangle=\sqrt{i\hbar\partial_t}\langle t|\phi(q)\rangle=\phi(t;q)$)
\begin{equation}
\label{DEpm}
     \left(\alpha\sqrt{2mi\hbar\partial_t}-\beta\sqrt{2m\mathcal{V}(q)} \right)\phi_{\mbox{\tiny L,R}}(t;q) = -i\hbar\partial_q\phi_{\mbox{\tiny L,R}}(t;q),
\end{equation}
where $\sqrt{\partial_t}$ is  a $1/2$-fractional time derivative \cite{Tarasov2012,BookLaskin2018}.  Applying $\sqrt{\partial_t}$ from the left on these equations we obtain the time-dependent SE: $i\hbar\partial_t \phi_{\mbox{\tiny L,R}}(t;q)=\mathcal{\hat H}_{\mbox{\tiny L,R}}^{\mbox{\tiny eff}}\phi_{\mbox{\tiny L,R}}(t;q)$ (we use $\sqrt{\partial_t}\sqrt{\partial_t}=\partial_t$ and assume $\sqrt{\partial_t}\partial_q=\partial_q\sqrt{\partial_t}$ \footnote{This is true when considering, for example, the $1/2$ Caputo-Liouville fractional derivative.}) with the effective Hamiltonian
\begin{equation}
\label{Veff}
    \mathcal{\hat H}_{\mbox{\tiny L,R}}^{\mbox{\tiny eff}}= -\frac{\hbar^2}{2m}\partial_q^2+
    \mathcal{V}(q) \pm \frac{\hbar}{{\sqrt{2m}}} \frac{d}{dq}\sqrt{\mathcal{V}(q)}.
\end{equation} 
This reproduces exactly the SUSY Hamiltonian (\ref{Hsusy}) when $W(q)=\sqrt{\mathcal{V}(q)}$. Thus, the Hamiltonian pairs $\mathcal{\hat H}_{\mbox{\tiny L,R}}^{\mbox{\tiny eff}}$ are {\it exactly} those given by SUSY, satisfying  the $\mathbb{Z}_2$ grading, namely  $\mathcal{H}_T=\mathcal{H}_{\mbox{\tiny R}}\oplus\mathcal{H}_{\mbox{\tiny L}}= \mathcal{H}^{\mbox{\tiny eff}}_{\mbox{\tiny R}}\oplus\mathcal{H}^{\mbox{\tiny eff}}_{\mbox{\tiny L}}$. 
Using separation of variables $\phi_{\mbox{\tiny L,R}}(t;q)= \psi_{\mbox{\tiny L,R}}(t)\chi_{\mbox{\tiny L,R}}(q)$, we obtain two coupled space-independent equations 
    \begin{eqnarray}
    \label{2T}
        \sqrt{2mi\hbar}\,D_t^{1/2}\psi_{\mbox{\tiny L,R}}(t)&=&\mathfrak{p}_{\mbox{\tiny L,R}}\psi_{\mbox{\tiny L,R}}(t),
    \end{eqnarray}
with $D_t^{1/2}=\sqrt{\partial_t}$ representing the $1/2$-fractional derivative in the Caputo sense \cite{Tarasov2012}, and separation constants $\mathfrak{p}_{\mbox{\tiny L,R}}$ having units of momentum. The two coupled time-independent SEs are
\begin{eqnarray}
\label{2Q}
    \left[\pm\frac{i\hbar}{\sqrt{2m}}\frac{d}{dq} {-}i\sqrt{\mathcal{V}(q)}\right] \chi_{\mbox{\tiny L,R}}(q)&=& \frac{\mathfrak{p}_{\mbox{\tiny R,L}}}{\sqrt{2m}}\,\chi_{\mbox{\tiny R,L}}(q).
\end{eqnarray}
Identifying $W(q)=\sqrt{\mathcal{V}(q)}$, the left-hand side of Eq.~(\ref{2Q}) naturally provide the SUSY operators (\ref{As})
and  supercharges (\ref{Qs}), obeying SUSY properties $\{\hat Q,\hat Q^{\dagger}\}=\mathcal{\hat H}_{\mbox{\tiny L,R}}^{\mbox{\tiny eff}}$, $[\hat Q,\hat Q]=[\hat Q^{\dagger},\hat Q^{\dagger}]=[\mathcal{\hat H}_{\mbox{\tiny L,R}}^{\mbox{\tiny eff}},\hat Q]=[\mathcal{\hat H}_{\mbox{\tiny L,R}}^{\mbox{\tiny eff}},\hat Q^{\dagger}]=0$. The SUSY algebra is realized without additional assumptions, emerging directly from the factorized Momentunian in the TSI QM.  By contrast to the standard SUSY, the superpotential is directly tied to the physical potential, reflecting the kinematic origin of SUSY.


Zero mode states are obtained using $\hat A\chi^0_{\mbox{\tiny R}}(q)=0$  and $\hat A^{\dagger}\chi^0_{\mbox{\tiny L}}(q)=0$, which corresponds to assume $\mathfrak{p}_{\mbox{\tiny L,R}}=0$ in {Eq.~(\ref{2Q}). Consequently, besides being zero-energy states, the usual focus in SUSY, these are also \textit{zero-momentum states}, which is in accordance to the definition of vacuum states in Quantum Field Theory (QFT)~\cite{weinberg1995quantum}. For vacuum states, Eqs.~(\ref{2T}) and~(\ref{2Q}) decouple, implying that the SUSY sectors become independent. In this limit, the solutions of the time-dependent SEs $\psi^0_{\mbox{\tiny L,R}}(t)$ are constant ($=C$) \footnote{The Caputo derivative of a constant is zero \cite{Tarasov2012}.}, and consequently the composite functions $\phi^0_{\mbox{\tiny L,R}}(t;q) = C\chi^0_{\mbox{\tiny L,R}}(q)$ become timeless.  If the states $\chi^0_{\mbox{\tiny L,R}}$ are normalizable, this situation is SUSY unbroken since $\ker Q \neq \varnothing$. Within the subspace $\mathcal{H}_T$, the zero-momentum states admit a clear physical interpretation as nonpropagating spatial configurations. The system occupies a configuration that cannot evolve further unless momentum is injected via $\mathfrak{p}_{\mbox{\tiny L,R}}\ne 0$. Thus, time evolution itself becomes a controlled phenomenon, activated only when the system is driven away from the protected sector.

As an example, for the harmonic oscillator (HO) the effective potential is
    $\mathcal{V}^{\mbox{\tiny HO,eff}}_{\mbox{\tiny L,R}}(q)=\frac12m\omega^2q^2\pm\frac{\hbar\omega}{2}\operatorname{sgn}{(q)}$. The associated SE is $\mathcal{\hat H}_{\mbox{\tiny L,R}}^{\mbox{\tiny HO,eff}}\chi_{\mbox{\tiny L,R}}(q) = \mathcal{K}^t\chi_{\mbox{\tiny L,R}}(q)$,
where $\mathcal{K}^t=\mathfrak{p}_{\mbox{\tiny\text{R}}}\mathfrak{p}_{\mbox{\tiny\text{L}}}/2m$ is the kinetic energy of the time, which is positive when $\mathfrak{p}_{\mbox{\tiny L,R}}$ have equal signs and negative otherwise, meaning respectively forward and backward time evolutions \cite{BeimsLara2024}. The additional term  ($\pm\frac{\hbar\omega}{2}$) in the effective potential can be transferred to the right-hand side of the above SE. After this, since the remaining terms on the left-hand side of the SE belong to the HO with energies  $E_n=(n+1/2)\hbar\omega$, it follows the quantization of the kinetic energy $\mathcal{K}^t = (n+1/2\mp\operatorname{sgn}(q)/2)\hbar\omega$, where $n \in \mathbb{N}$. Consequently, the quantity $p_n \equiv \pm\sqrt{\mathfrak{p}_{\mbox{\tiny R}}\mathfrak{p}_{\mbox{\tiny L}}} = \pm \sqrt{ \operatorname{sgn}(\mathfrak{p}_{\mbox{\tiny R}}\mathfrak{p}_{\mbox{\tiny L}})} \sqrt{2m (n+1/2\mp\operatorname{sgn}(q)/2)\hbar\omega}$ is also quantized. All pairs of excited states can be obtained using $\mathfrak{p}_{\mbox{\tiny L,R}}\ne 0$, leading to broken SUSY. The set $\{p_n \}$ constitute the spectrum of $\hat{\cal P}^{\pm}(\hat{\mathcal{H}},\hat t;q)$ for the HO. The solutions for the zero-momentum $(\mathfrak{p}_{\mbox{\tiny L,R}}=0)$ ground states ($p_{n}=0 \Rightarrow n=0,-1$) are
    $\chi^0_{\mbox{\tiny L,R}}(q)= C_{\mbox{\tiny L,R}}\exp\!\left(\mp\frac{m\omega}{2\hbar}\,q^{2}\right)$, which are identical to the zero-energy mode states from SUSY.  One sector ($n=0$) admits a normalizable zero-momentum ground state $\chi^0_{\mbox{\tiny R}}(q)$, while the corresponding state in the partner sector ($n=-1$) is non-normalizable and therefore excluded from the physical spectrum. In this case the Witten index is $\Delta=1$, being SUSY unbroken.

\noindent
 \textit{Relativistic Momentunian.} 
Following an analogous classical derivation described in \cite{BeimsLara2024}, for the non-relativistic case we obtain 
\begin{equation}
    {\cal P}^{\pm}(\mathcal{H},t,q)    = \pm \frac{1}{c}\sqrt{\left[\mathcal{H} - \mathcal{V}(t,q)\right]^2 - (m_0 c^2)^2},
\end{equation}
which matches the relativistic momentum $p_q = m_0 \gamma v$ ($v=dq/dt$), with $\gamma(v) = 1/\sqrt{1 - \beta^2}$ and $\beta = v/c$, and  the signal indicating direction of movement in the $q-$axis. 

Promoting $t$ and $\mathcal{H}$ to operators and applying the Dirac method we get $\hat{\cal P}({\hat{\mathcal{H}},\hat t;q}) = \alpha \frac{1}{c} \left[\hat{\mathcal{H}} - \mathcal{V}(\hat t,q) \right] + \beta m_0 c$. Using Pauli matrices $\alpha = \sigma_x$ and $\beta = i\sigma_z$, the corresponding Dirac-like equations become
\begin{equation}
    \frac{1}{c}\left[ i \hbar\,\partial_t  -\mathcal{V}(t,q)\right]\phi_{\mbox{\tiny L,R}}(t;q) = \left[  - i \hbar\,\partial_q \pm i\, m_0 c\right] \phi_{\mbox{\tiny R,L}}(t;q).\label{phiRel}
    \nonumber
\end{equation}
 
To proceed we use separation of variables $\phi_{\mbox{\tiny L,R}}(t,q)=\chi_{\mbox{\tiny L,R}}(q)\psi_{\mbox{\tiny L,R}}(t)$. Assuming $\mathcal{V}(t,q)=\mathcal{V}(t)$ we obtain $\psi_{\mbox{\tiny L}}(t)= \psi_{\mbox{\tiny R}}(t)=e^{-\frac{i}{\hbar}\left(c\,\mathfrak{p}t +\int^t \mathcal{V}(\tau)\tau\right)}$ and
\begin{equation}
    \left(-i \hbar\,\frac{d}{dq}\pm {i} m_0 c\right) \chi_{\mbox{\tiny R,L}}= \mathfrak{p}\,\chi_{\mbox{\tiny L,R}} \label{pc}, 
\end{equation}
which leads to the usual relativistic Dirac equation with the separation constant $\mathfrak{p}$ having units of momentum. For $\mathcal{V}(t,q)=\mathcal{V}(q)$ we obtain $\psi_{\mbox{\tiny L}}(t)= \psi_{\mbox{\tiny R}}(t)=e^{-\frac{i}{\hbar}c\,\mathfrak{p}t}$ and
\begin{equation}
\label{pVc}
    \left[  - i \hbar\,\frac{d}{dq} \pm {i} m_0 c\right] \chi_{\mbox{\tiny R,L}}(q)= \left(\mathfrak{p}-\frac{\mathcal{V}(q)}{c}\right)\chi_{\mbox{\tiny L,R}}(q).
\end{equation}
To eliminate first-orders derivatives we use $\chi_{\mbox{\tiny L,R}}(q)=\sqrt{R(q)}\,\eta_{\mbox{\tiny L,R}}(q)$ 
with $R(q)=\mathfrak{p}-\mathcal{V}(q)/c$, and after straightforward calculation we obtain the SE:
$\mathcal{\hat H}_{\mbox{\tiny L,R}}^{\mathrm{eff}}(q)\eta_{\mbox{\tiny L,R}}(q)
= -\frac{m_0c^2}{2}\eta_{\mbox{\tiny L,R}}(q),$
with Hamiltonian partners $\mathcal{\hat H}_{\mbox{\tiny L,R}}^{\mathrm{eff}}(q)=\frac{\hbar^2}{2m_0} \frac{d^2}{dq^2}+V_{\mbox{\tiny L,R}}^{\mathrm{eff}}(q)$
and effective potential
\begin{eqnarray}
    V_{\mbox{\tiny L,R}}^{\mathrm{eff}}(q)= \frac{3\hbar^2 (R')^2}{8m_0R^2} - \frac{\hbar^2 R''}{4m_0R} -\frac{R^2}{2m_0}  \pm \hbar c \frac{R'}{2R}.
\end{eqnarray}

From Eqs.~(\ref{pc}) and (\ref{pVc}) we identify the generators $A=- i \hbar\,\frac{d}{dq}+ {i} m_0c$ and $A^{\dagger}=- i \hbar\,\frac{d}{dq}-{i} m_0c$, independent of the potential. Using the definitions (\ref{Qs}) and $P^2=\{\hat Q,\hat Q^{\dagger}\}$ we obtain $P^2=H^2_{\mbox{\tiny D}}/c^2$, where $H_{\mbox{\tiny D}}$ is the usual 1D Dirac Hamiltonian for the free particle. In the case of Eq.~(\ref{pVc}), the effect of the potential comes through the constraint $R(q)=0$ to obtain the zero-mode states, namely $\mathfrak{p}=\mathcal{V}(q)/c$. This is analogous to apply $A\,\chi^0_{\mbox{\tiny L}}=0$ and $A^{\dagger}\chi^0_{\mbox{\tiny R}}=0$ and leads to
 \begin{equation}
 \label{JJ}
    \phi_{\mbox{\tiny L,R}}^0(t,q)= e^{-\frac{i}{\hbar}Et}\, e^{\mp\frac{1}{\hbar}m_0c\,q},
\end{equation}
where we identified $\mathfrak{p}=E/c$. The quantity $E-\mathcal{V}(q)$ is the kinetic-part of the relativistic available energy, which for a massive particle must be $\ge m_0c^2$. Thus, the constraint $E-\mathcal{V}(q)=0$ forces an imaginary momentum and the evanescent relativistic solution (\ref{JJ}). These solutions have the same form as Jackiw–Rebbi zero modes states \cite{Jackiw1976Solitons} if we introduce a position-dependent mass $m(q)=m_0\operatorname{sgn}(q)$, and is a way to  guarantee normalization.

\noindent
\textit{$\sqrt{\mbox{T}}$-SUSY: Non-relativistic Momentunian partners.}
\label{Mpartnes}
Since the TSI QM naturally incorporates SUSY-like operators, it is interesting to check the existence and relevance of Momentunian partners. To factorize each Momentunian $\sqrt{\mbox{T}}$-SUSY sector we use the auxiliary function $W(q)$ to define
\begin{subequations}
    \begin{eqnarray}
        \hat A &=& (2m)^{1/4}\sqrt{  \sqrt{\mathcal{\hat H}} + \sqrt{ W(q)}}, \\
        \hat B &=& (2m)^{1/4}\sqrt{  \sqrt{\mathcal{\hat H}} - \sqrt{ W(q)}},
    \end{eqnarray}
\end{subequations}
and supercharges (\ref{Qs}). In this way we obtain $\mathcal{\hat P}^{\mbox{\tiny (\textpm)}}_{\mbox{\tiny L,R}}= \sigma_z \{\hat Q,\hat Q^{\#} \}=\begin{pmatrix}\hat A\hat B&0\\0&\hat B\hat A\end{pmatrix}$, and the SUSY algebra is satisfied. Since $[\hat A,\hat B]=0$, we have $[\mathcal{\hat P}^{\mbox{\tiny (\textpm)}}_{\mbox{\tiny L,R}},\hat Q]=[\mathcal{\hat P}^{\mbox{\tiny (\textpm)}}_{\mbox{\tiny L,R}},\hat Q^{\#}]=0$. Using $\hat B\ne \hat A^{\dagger}$, our construction is analogous—apart from the presence of fractional supercharges—to a non-Hermitian realization of SUSY \cite{Bagarello2020}. Rewritten compactly we have
 $\mathcal{\hat P}^{\mbox{\tiny (\textpm)}}_{\mbox{\tiny L,R}}= \sigma_z\sqrt{2m\left(\mathcal{\hat H} -  W(q)  \right)}$,
so that $\mathcal{\hat P}^{\mbox{\tiny (\textpm)}}_{\mbox{\tiny L}}= \mathcal{\hat P}^{\mbox{\tiny (\textpm)}}_{\mbox{\tiny R}}$. The zero-momentum mode states are obtained from $\hat A\phi_{\mbox{\tiny R}}^0(t;q)=0$  and $\hat B\phi_{\mbox{\tiny L}}^0(t;q)=0$,  leading to 
\begin{equation}
    \label{phi0F1}
    \sqrt{i\hbar\partial_t}\,\phi_{\mbox{\tiny L,R}}^0(t,q)=\mp\sqrt{W(q)}\,\phi_{\mbox{\tiny L,R}}^0(t,q).
\end{equation}
Applying $\partial_t^{1/2}$ from the left we obtain the SE-like: 
%
    $i\hbar\partial_t\,\phi_{\mbox{\tiny L,R}}^0(t,q)=W(q)\,\phi_{\mbox{\tiny L,R}}^0(t,q)$,
%
showing that only the superpotential is responsible for the time evolution of the zero mode states. As before, the contribution of the kinetic part is absent, compatible with the zero-momentum state. The solution of the fractional differential Eq.~(\ref{phi0F1}) is given in terms of the Mittag--Leffler function, namely \cite{Lim2012}
\begin{equation}
    \phi_{\mbox{\tiny L,R}}^0(t,q)=\phi_{\mbox{\tiny L,R}}^0(q)\,E_{1/2}\!\left(\mp\frac{\sqrt{W(q)}}{\sqrt{i\hbar}}\sqrt{t}\right).
\end{equation}
In distinction to the positivity of the Hamiltonian in the usual SUSY,  the Momentunian can have positive and negatives values, due to a symmetry. Defining the symmetry operator $ S= \sigma_x$ and using the Pauli algebra $\sigma_x \sigma_z \sigma_x = -\sigma_z$, we obtain $ S\mathcal{\hat P}^{\mbox{\tiny (\textpm)}}_{\mbox{\tiny L,R}}S^{-1} = -\mathcal{\hat P}^{\mbox{\tiny (\textpm)}}_{\mbox{\tiny L,R}}$. Assume that $|\phi\rangle$ is an eigenstate of the momentunian with eigenvalues $\pm\,\mathfrak{p}$.  Acting with $S$ gives $\mathcal{\hat P}^{\mbox{\tiny (\textpm)}}_{\mbox{\tiny L,R}} (S|\phi\rangle)= \mp S \mathcal{\hat P}^{\mbox{\tiny (\textpm)}}_{\mbox{\tiny L,R}} |\phi\rangle= \mp\mathfrak{p} (S|\phi\rangle)$.  Thus, the symmetry enforces the pairing $\mathfrak{p} \longleftrightarrow -\mathfrak{p}$, and at the point $\mathfrak{p} = 0$ the symmetry maps the state into itself.  Consequently, the zero-momentum subspace naturally splits into two symmetric sectors. Considering a perturbation that preserves the symmetry, $S \mathcal{\hat P}^{\mbox{\tiny (\textpm)} \prime}_{\mbox{\tiny L,R}} S^{-1} = \mp\mathcal{\hat P}^{\mbox{\tiny (\textpm)} \prime}_{\mbox{\tiny L,R}}$, the spectrum must remain symmetric. Therefore, an isolated state at $\mathfrak{p}=0$ cannot move away from zero without creating a pair of states with eigenvalues $\pm \mathfrak{p}$. Such mechanism is analogous to the protection of Dirac nodes in systems described by a Dirac Hamiltonian $H = v_F (\sigma_x p_x + \sigma_y p_y)$ where a chiral symmetry implies $\{\sigma_z,H\} = 0$ \cite{CastroNeto2009}. In that case the symmetry enforces $E \leftrightarrow -E$, so that the point $E=0$ becomes protected. 

\noindent
\textit{$\sqrt{\mbox{T}}$-SUSY: The relativistic Momentunian partners.}
\label{subsec:RMP}
Finally, we factorize each relativistic Momentunian sector by defining the operators
\begin{subequations}
    \begin{eqnarray}
        \hat A &=& \sqrt{\frac{1}{c}\left( \mathcal{\hat H} - W(q) +m_0c^2 \right)},\\
        \hat B &=& \sqrt{\frac{1}{c}\left( \mathcal{\hat H} + W(q)-m_0c^2 \right)},
    \end{eqnarray}
\end{subequations}
and using the supercharges (\ref{Qs}) we obtain
%
    $\hat{\mathcal{P}}^{(\pm)}_{\mbox{\tiny L,R}} =\frac{1}{c}\sqrt{\left(\mathcal{\hat H}^2- W^2(q)\;- (m_0c^2)^2\right)}$.
The zero mode states are obtained from $\hat Q\phi^0_{\mbox{\tiny R}}(t;q)=0$ and $\hat Q^{\#}\phi^0_{\mbox{\tiny L}}(t;q)=0$ leading to:
%
    $i\hbar\partial_t\,\phi^0_{\mbox{\tiny L,R}}(t,q)=\pm (W(q)+m_0c^2 )\,\phi^0_{\mbox{\tiny L,R}}(t,q)$,
%
and have the formal solution
\begin{equation}
    \phi^0_{\mbox{\tiny L,R}}(q,t)=\phi^0_{\mbox{\tiny L,R}}(q)\,\exp\!\left[\pm\frac{i}{\hbar}\left( m_0c^2\,t +W(q)\,t\right)\right].
    \end{equation}
The zero-momentum-mode states are essentially stationary states in the rest frame of the particle. In relativistic mechanics, the rest frame is defined by vanishing momentum, and the total energy reduces to $E = m_0 c^2 + W(q)$, leading to the time-dependent phase of the state  $\exp\!\left[\pm \frac{i}{\hbar} E\,t \right]$, which corresponds to the evolution of a state with zero kinetic energy. For the free particle $W(q)=0$ the time-constraint here forces the system to evolve at a frequency $\omega = m_0c^2/\hbar$, proportional to its invariant rest mass.

\noindent
\textit{Conclusions.}
In this Letter, we demonstrated that supersymmetry emerges naturally within the framework of TSI QM. By factorizing the Momentunian operator, we obtained a supersymmetric algebra with well-defined supercharges and partner Hamiltonians, without introducing supersymmetry phenomenologically. In the non-relativistic formulation, the superpotential is directly connected to the physical potential, and the spectrum can also be organized in momentum doublets rather than the usual energy doublets. The formalism also naturally incorporates $1/2$-fractional time derivatives, linking supersymmetry with memory effects in the temporal evolution.
For the relativistic case, the zero-mode solutions correspond to evanescent states. We further introduced non-relativistic and relativistic Momentunian partners with non-Hermitian supercharges, defining what we call $\sqrt{T}$-SUSY. Altogether, these results indicate that supersymmetry may be an intrinsic structural property of the TSI quantum dynamics, opening new perspectives on the relation between temporal observables, relativistic dynamics, memory effects and supersymmetric QM.

\vspace*{-0.5cm}
\section*{Acknowledgements}
The Authors thank Profs.~I Kuntz and JM Rost for helpful discussions. MWB thanks the National Council for Scientific and Technological Development—CNPq (Brazilian agency) for financial support (Grant Number. 310294/2022-3).

\section*{Author Contributions}
MWB conceptualized this work. MWB and AJSL contributed equally to this work.

\end{document}